\documentclass[aps,prd,reprint,showpacs,showkeys,superscriptaddress]{revtex4-1}
\usepackage{amsmath}
\usepackage{amssymb}
\usepackage{bm}
\usepackage{graphicx}
\usepackage[pdftex,colorlinks,hyperindex,plainpages=false,bookmarksopen,bookmarksnumbered,pdfusetitle]{hyperref}
\usepackage{mathrsfs}
\usepackage{slashed}

\let\de=\partial
\newcommand\dd{\text{d}}
\newcommand\imag{\text{i}}
\newcommand\La{\mathscr{L}}
\newcommand\De{\mathscr{D}}
\newcommand\muI{\mu_\text{I}}
\newcommand\nI{n_\text{I}}
\newcommand\gr[1]{\text{#1}}
\newcommand\he[1]{#1^\dagger}
\newcommand\vek[1]{\bm{#1}}
\DeclareMathOperator{\Tr}{Tr}
\DeclareMathOperator{\tr}{tr}

\begin{document}

\title{Vector meson condensation in a pion superfluid}

\author{Tom\'{a}\v{s} Brauner}
\email{tomas.brauner@uis.no}
\affiliation{Faculty of Science and Technology, University of Stavanger, 4036 Stavanger, Norway}
\author{Xu-Guang Huang}
\email{huangxuguang@fudan.edu.cn}
\affiliation{Physics Department and Center for Particle Physics and Field Theory, Fudan University, Shanghai 200433, China}

\begin{abstract}
We revisit the suggestion that charged $\rho$-mesons undergo Bose-Einstein condensation in isospin-rich nuclear matter. Using a simple version of the Nambu-Jona-Lasinio (NJL) model, we conclude that $\rho$-meson condensation is either avoided or postponed to isospin chemical potentials much higher than the $\rho$-meson mass as a consequence of the repulsive interaction with the preformed pion condensate. In order to support our numerical results, we work out a linear sigma model for pions and $\rho$-mesons, showing that the two models lead to similar patterns of medium dependence of meson masses. As a byproduct, we analyze in detail the mapping between the NJL model and the linear sigma model, focusing on conditions that must be satisfied for a quantitative agreement between the models.
\end{abstract}

\maketitle


\section{Introduction}
\label{sec:intro}

The phase structure of nuclear and quark matter at low temperatures and high densities remains one of the major unresolved problems in contemporary physics. While direct Monte Carlo simulations of Quantum ChromoDynamics (QCD) at high baryon density are inhibited by the notorious sign problem, this does not affect QCD at nonzero \emph{isospin} density~\cite{Son:2000xc}. Although the latter is not entirely physical, since a medium with nonzero isospin but zero net baryon density does not seem to exist in nature, the study of QCD under such conditions may shed some light on ordinary nuclear matter.

The fact that the lightest hadron carrying isospin, the pion, is a pseudo-Nambu-Goldstone (NG) boson of spontaneously broken chiral symmetry, allows to draw rigorous conclusions about the QCD phase diagram at nonzero isospin density. At zero temperature, charged pions undergo Bose-Einstein condensation (BEC) once the isospin chemical potential $\muI$ exceeds the pion mass, $m_\pi$~\footnote{This statement is rigorous for QCD with nonzero isospin chemical potential. However, there is evidence that pion BEC occurs also in a hadron medium at nonzero baryon density~\cite{Voskresensky:1994uz,*Ebert:2005cs}.}. In the opposite extreme of very high isospin chemical potential, one expects a crossover from meson-dominated to quark-dominated matter. Such high-isospin-density matter is characterized by a first-order confinement-deconfinement transition at temperatures much lower than the QCD scale~\cite{Son:2000xc,Cohen:2015soa}. What happens at intermediate isospin densities remains unclear though.

Since the next-to-lightest particle in terms of the mass-to-isospin ratio is the $\rho$-meson, it was suggested early on that at sufficiently high $\muI$, charged $\rho$-mesons will undergo BEC as well~\cite{Voskresensky:1997ub,*Lenaghan:2001sd,*Sannino:2002wp}. This would have far-reaching consequences for the structure of isospin-rich nuclear matter as it would, among others, imply spontaneous breaking of rotational invariance. However, the fate of $\rho$-mesons in the pion superfluid phase is not that obvious, since isospin symmetry is spontaneously broken in this phase and hence the $\rho$-meson mass no longer depends linearly on the chemical potential. Unfortunately, the issue cannot be settled within the model-independent approach of chiral perturbation theory ($\chi$PT)~\cite{Gasser:1983yg} since it addresses physics outside of $\chi$PT's range of validity. A holographic model for QCD at nonzero isospin density was proposed in Ref.~\cite{Aharony:2007uu}, concluding that, indeed, $\rho$-mesons condense for sufficiently high values of $\muI$ ($\muI\gtrsim1.7m_\rho$ in their setup). This work is, nevertheless, based on QCD in the large-$N_c$ limit and in the chiral limit, where the theory possesses a continuous axial $\gr{U(1)}$ symmetry.

A preliminary lattice study of QCD with two flavors of Wilson fermions with focus on the meson spectrum was carried out in Ref.~\cite{forcrandlattice}. It was found that at sufficiently high values of $\muI$, the charged $\rho$-meson mass drops, within the numerical accuracy, to zero, which suggests BEC. However, their results are at odds with other model-independent predictions of $\chi$PT. The latter implies that within the pion superfluid phase, the mass of the neutral pion equals $\muI$. In fact, this is now understood to be an \emph{exact} consequence of spontaneous breaking of isospin by the charged pion condensate: the neutral pion is the massive NG boson of the isospin symmetry~\cite{Watanabe:2013uya}. This is in contrast to the findings of Ref.~\cite{forcrandlattice}, according to which the neutral pion mass starts to decrease with chemical potential upon the onset of BEC at $\muI=m_\pi$.

Motivated by these works, the objective of the present paper is to provide new insight into the question whether $\rho$-mesons condense at sufficiently high isospin chemical potential. Our primary tool is the Nambu-Jona-Lasinio (NJL) model. Being merely a model, this, of course, does not allow us to make \emph{rigorous} conclusions about the QCD phase diagram. Still, the NJL model is known to work reasonably for vacuum physics~\cite{Vogl:1991qt,*Klevansky:1992qe} and its advantage is a relatively low number of free parameters, all of which can be fixed by a fit to vacuum observables. The analysis is carried out in Sec.~\ref{sec:NJL}, starting with a detailed discussion of the condensates in the pion superfluid phase, and proceeding to the meson spectrum therein. Treating the vacuum $\rho$-meson mass as a free parameter, we show that if present at all, $\rho$-meson BEC is postponed to chemical potentials far beyond the scope of the NJL model.

In order to gain deeper insight into the nature of this result, we complement the NJL analysis by one using the linear sigma model in Sec.~\ref{sec:lsm}. It turns out that the repulsive interaction between pions and $\rho$-mesons results in postponing the onset of $\rho$-meson BEC. While the latter always occurs in the linear sigma model, for arbitrary model parameters and sufficiently high $\muI$, it can indeed be pushed to values $\muI\gg m_\rho$.

Apart from our main results, some aspects of the presented work might be useful in a broader context for those working on effective model description of quark matter. First, to the best of our knowledge, this is the first time that the condensates of the temporal vector and axial vector field in the pion superfluid phase have been calculated self-consistently using the NJL model. We show how the results obtained in the NJL model match the predictions of $\chi$PT. For the reader's convenience, the $\chi$PT analysis is summarized in Appendix~\ref{app:ChPT}. Second, we undertake a systematic study of the mapping between the NJL model and the linear sigma model. To what extent the predictions of the NJL model and the matched linear sigma model agree is discussed in Appendix~\ref{app:matching}.


\section{Nambu-Jona-Lasinio model analysis}
\label{sec:NJL}

The basic degrees of freedom of the NJL model are the quarks. Their detailed dynamics depends on the choice of interaction which is to some extent arbitrary. We want to take advantage of the simplicity of the mean-field approximation, hence we must include interaction channels with the quantum numbers of the pions and $\rho$-mesons. A minimal NJL-type model that fulfills this requirement and at the same time respects the full chiral symmetry of QCD is the two-flavor model defined by the Lagrangian
\begin{equation}
\begin{split}
\La={}&\bar\psi(\imag\slashed D-m)\psi+G\bigl[(\bar\psi\psi)^2+(\bar\psi\imag\gamma_5\vec\tau\psi)^2\bigr]\\
&-G_V\bigl[(\bar\psi\gamma_\mu\vec\tau\psi)^2+(\bar\psi\gamma_\mu\gamma_5\vec\tau\psi)^2\bigr].
\end{split}
\label{NJLlag}
\end{equation}
(A similar model was used in Ref.~\cite{Chernodub:2011mc} to study $\rho$-meson condensation in the QCD vacuum in a magnetic field.) Here $\psi$ is the isospin-doublet quark field and $\vec\tau$ the Pauli matrices in isospin space. The covariant derivative includes the isospin chemical potential via $D_0\equiv\de_0-\imag\muI\frac{\tau_3}2$. The model is defined completely by four parameters: the current quark mass $m$ (assumed for simplicity to be the same for both quark flavors), the scalar and vector coupling $G$ and $G_V$, and the ultraviolet cutoff $\Lambda$.

We follow the usual bosonization procedure wherein the four-point interaction is decoupled by introducing a set of collective bosonic fields, $\sigma$, $\vec\pi$, $\vec\rho_\mu$ and $\vec a_\mu$, and subsequently carrying out the Hubbard-Stratonovich transformation. This brings the Lagrangian~\eqref{NJLlag} to the form
\begin{equation}
\La=-\frac{\sigma^2+\vec\pi^2}{4G}+\frac{\vec\rho_\mu^2+\vec a_\mu^2}{4G_V}+\bar\psi\imag\De\psi,
\label{NJLsemibosonized}
\end{equation}
where the modified Dirac operator of the theory reads
\begin{equation}
\imag\De\equiv\imag\slashed D-m-\sigma-\imag\gamma_5\vec\pi\cdot\vec\tau+\vec{\slashed\rho}\cdot\vec\tau+\vec{\slashed a}\cdot\vec\tau\gamma_5.
\label{diracoperator}
\end{equation}
Integrating out the quark degrees of freedom then leads to the fully bosonized effective action
\begin{equation}
S_\text{eff}=\int\dd^4x\left(-\frac{\sigma^2+\vec\pi^2}{4G}+\frac{\vec\rho_\mu^2+\vec a_\mu^2}{4G_V}\right)-\imag\Tr\log\De.
\label{NJLeffaction}
\end{equation}
This effective action is the starting point for the calculation of both the thermodynamic potential (and hence the phase structure of the model) and the meson spectrum.


\subsection{Vacuum physics}
\label{subsec:NJLvac}

In the QCD vacuum, isospin is not spontaneously broken~\cite{Vafa:1983tf}. Hence only the $\sigma$ field can acquire a vacuum expectation value, which is found self-consistently from the gap equation $\delta S_\text{eff}/\delta\sigma=0$. After a simple manipulation, the gap equation can be brought to the form
\begin{equation}
\langle\sigma\rangle=16GN_cMI_1,
\label{gapeqsigma}
\end{equation}
where $I_1$ denotes the tadpole momentum integral,
\begin{equation}
I_1\equiv\imag\int\frac{\dd^4k}{(2\pi)^4}\frac1{k^2-M^2},
\label{tadpole}
\end{equation}
and we explicitly highlighted the dependence on the number of colors, $N_c=3$. In both expressions, $M\equiv m+\langle\sigma\rangle$ denotes the dynamical (constituent) quark mass. While the vacuum expectation value of $\sigma$ is an observable specific to the present NJL model, it can be directly related to the expectation value of the scalar operator composed of the quark fields,
\begin{equation}
\langle\bar\psi\psi\rangle=-\frac{\delta S_\text{eff}}{\delta m}=-\frac{\langle\sigma\rangle}{2G},
\label{NJLchiralcond}
\end{equation}
where the gap equation for $\langle\sigma\rangle$ has already been used. This, together with Eq.~\eqref{gapeqsigma}, constitutes one of the matching relations used to fix our model parameters.

The other observables used to fix the model parameters are related to the meson spectrum in the vacuum. This can be extracted from the meson polarization functions, defined by the second functional derivative of the effective action~\eqref{NJLeffaction}, symbolically
\begin{equation}
\chi^{(\mathcal{AB})}(x-y)\equiv\frac{\delta^2S_\text{eff}}{\delta\phi_\mathcal{A}(x)\delta\phi_\mathcal{B}(y)},
\label{NJLpolarization}
\end{equation}
where $\phi_\mathcal{A}$ runs over all the meson fields of the model and the multi-index $\mathcal A$ includes the field type as well as the isospin and Lorentz indices. As a consequence of conservation of isospin and parity~\cite{Vafa:1984xg,*Azcoiti:1999rq,*Ji:2001sa} in the QCD vacuum, the polarization function matrix will take a block-diagonal form. The simplest block is that of the sole isospin singlet of the model, $\sigma$. We will not discuss it in detail here as it is of no interest for our purposes though.

The other meson modes are all isospin triplets. The polarization function must clearly be diagonal in the isospin space. Denoting isospin indices as $a,b$, it can be written as $\chi^{(AB)}_{ab}(p)\equiv\chi^{(AB)}(p)\delta_{ab}$ upon Fourier transform to momentum space, where the multi-index $A$ now only labels the field type and the Lorentz index. The polarization functions $\chi^{(AB)}$ can in turn be obtained by evaluating the right-hand side of Eq.~\eqref{NJLpolarization}, which is equivalent to a one-loop integral with two external meson legs,
\begin{equation}
\begin{split}
\chi^{(AB)}(p)={}&2G^{(AB)}+2\imag N_c\int\frac{\dd^4k}{(2\pi)^4}\\
&\times\frac{\tr_D\bigl[\Gamma^{(A)}(\slashed k+\slashed p+M)\Gamma^{(B)}(\slashed k+M)\bigr]}{\bigl[(k+p)^2-M^2\bigr](k^2-M^2)}.
\end{split}
\label{chiisosinglet}
\end{equation}
Here $G^{(AB)}$ is a constant diagonal matrix defined by $G^{(\pi\pi)}=-1/(4G)$ and $G^{(\rho\rho)}_{\mu\nu}=G^{(aa)}_{\mu\nu}=g_{\mu\nu}/(4G_V)$. Also, the matrices $\Gamma^{(A)}$ carry information about quantum numbers of the modes, and read $\Gamma^{(\pi)}=\imag\gamma_5$, $\Gamma^{(\rho)}_\mu=-\gamma_\mu$ and $\Gamma^{(a)}_\mu=-\gamma_\mu\gamma_5$. The trace in Eq.~\eqref{chiisosinglet} is to be carried out over the space of Dirac matrices only. Upon some algebraic manipulation, all the polarization functions can be expressed in terms of two basic momentum integrals: the tadpole $I_1$~\eqref{tadpole} and the one-loop integral
\begin{equation}
I_2(p^2)\equiv-\imag\int\frac{\dd^4k}{(2\pi)^4}\frac1{\bigl[(k+p)^2-M^2\bigr](k^2-M^2)}.
\label{NJLIPiIntegral}
\end{equation}
Since we are going to regulate the divergent integrals using a sharp three-momentum cutoff, we give here also suitable expressions for both integrals after the frequency integration has been carried out,
\begin{equation}
\begin{split}
I_1&=\int\frac{\dd^3\vek k}{(2\pi)^3}\frac1{2\epsilon_{\vek k}},\\
I_2(p^2)&=\int\frac{\dd^3\vek k}{(2\pi)^3}\frac1{\epsilon_{\vek k}(4\epsilon_{\vek k}^2-p^2)},
\end{split}
\end{equation}
where $\epsilon_{\vek k}\equiv\sqrt{\vek k^2+M^2}$ is the fermion quasiparticle dispersion relation in the vacuum.

We next focus on the $\rho$-meson polarization function. Being a symmetric rank-two tensor, this can be decomposed into a longitudinal and transverse part,
\begin{equation}
\chi^{(\rho\rho)}_{\mu\nu}(p)\equiv\chi^L_{(\rho)}(p^2)\frac{p_\mu p_\nu}{p^2}+\chi^T_{(\rho)}(p^2)\left(g_{\mu\nu}-\frac{p_\mu p_\nu}{p^2}\right).
\label{longtrans}
\end{equation}
The longitudinal piece is constant, $\chi^L_{(\rho)}(p^2)=1/(2G_V)$, reflecting the fact that the longitudinal component of the field $\vec\rho_\mu$ does not describe a propagating degree of freedom. The three physical degrees of freedom of the massive vector meson are all contained in the transverse part,
\begin{equation}
\chi^T_{(\rho)}(p^2)=\frac1{2G_V}-\frac{16}3N_cI_1-\frac83N_c(p^2+2M^2)I_2(p^2).
\label{NJLrhopolar}
\end{equation}
This equation can be used to determine the $\rho$-meson mass in the vacuum once all the parameters of the model are known.

The case of the pion and the axial vector meson is somewhat more complicated~\cite{Klevansky:1997dk,He:1997gn}. Namely, the pion and the longitudinal component of the $\vec a_\mu$ field carry the same quantum numbers and thus can mix. The polarization function in this sector forms a nontrivial $2\times2$ matrix, represented by the components
\begin{equation}
\begin{split}
\chi^{(\pi\pi)}(p)&=-\frac1{2G}+8N_cI_1+4N_cp^2I_2(p^2),\\
\chi^{(\pi a)}_\mu(p)&=8\imag p_\mu N_cMI_2(p^2),\\
\chi^L_{(a)}(p^2)&=\frac1{2G_V}+16N_cM^2I_2(p^2).
\end{split}
\label{piamixing}
\end{equation}
The transverse part of $\vec a_\mu$ is characterized by the polarization function, defined similarly to Eq.~\eqref{longtrans},
\begin{equation}
\chi^T_{(a)}(p^2)=\frac1{2G_V}-\frac{16}3N_cI_1-\frac83N_c(p^2-4M^2)I_2(p^2),
\end{equation}
which in principle allows to determine the mass of the axial vector meson (the $a_1$-meson). Note that the transverse and longitudinal polarization functions of the vector and axial vector meson satisfy the sum rule~\cite{Klevansky:1997dk}
\begin{equation}
\chi^T_{(a)}(p^2)-\chi^T_{(\rho)}(p^2)=\chi^L_{(a)}(p^2)-\chi^L_{(\rho)}(p^2).
\end{equation}

We are now ready to put together the set of equations that can be used to fix the parameters of our model. As already mentioned above, there are altogether four parameters: $m$, $G$, $G_V$ and the cutoff $\Lambda$. We therefore need four observables on input, and it is most common to choose three of these as the chiral condensate $\langle\bar\psi\psi\rangle$, pion mass $m_\pi$ and decay constant $f_\pi$. The fourth one is naturally provided by the $\rho$-meson mass $m_\rho$. However, for the sake of convenience, we shall trade the chiral condensate for the constituent quark mass $M$. When combined with $m_\pi$ and $m_\rho$, this has the advantage of giving us a direct control over the threshold for the (unphysical) decay of the $\rho$-meson into a quark-antiquark pair, $\rho\to q\bar q$, as well as over the threshold for the (physical but absent at mean-field level) two-pion decay, $\rho\to\pi\pi$. All four equations needed for parameter fixing can thus be put together in the following compact form,
\begin{widetext}
\begin{equation}
\begin{alignedat}{4}
\langle\sigma\rangle&=16GN_cMI_1\qquad\qquad&&\text{(chiral condensate),}\\
\frac m{2GM}&=\frac{4N_cm_\pi^2I_2(m_\pi^2)}{1+32G_VN_cM^2I_2(m_\pi^2)}\qquad\qquad&&\text{(pion mass)},\\
f_\pi^2&=\frac{m\langle\sigma\rangle}{2Gm_\pi^2}\qquad\qquad&&\text{(pion decay constant)},\\
\frac1{2G_V}-\frac{\langle\sigma\rangle}{3GM}&=\frac83N_c(m_\rho^2+2M^2)I_2(m_\rho^2)\qquad\qquad&&\text{($\rho$-meson mass)}.
\end{alignedat}
\label{fixing}
\end{equation}
\end{widetext}
The condition for the pion mass was obtained from the mixing matrix~\eqref{piamixing} by integrating out the longitudinal component of $\vec a_\mu$, and setting the resulting effective pion polarization function to zero. Also, both here and in the condition for the $\rho$-meson mass, the gap equation~\eqref{gapeqsigma} was used to simplify the expressions. This makes it clear that the pion becomes massless in the chiral limit where $m\to0$. Finally, we note that the condition for the pion decay constant is not a result of an actual NJL model calculation; for the sake of simplicity, we used instead the Gell-Mann-Oakes-Renner relation~\cite{Klevansky:1992qe} rewritten in terms of $\langle\sigma\rangle$ using Eq.~\eqref{NJLchiralcond}. This is expected to give an accurate approximation to $f_\pi$ as long as the pion mass is well below the characteristic scale of spontaneous chiral symmetry breaking in the QCD vacuum.

As we show below, the tendency towards $\rho$-meson BEC depends very sensitively on the vacuum mass $m_\rho$. Taking its physical value, $m_\rho\approx775\text{ MeV}$, would require us to go to a range of $\muI$ way beyond the scope of the NJL model. We therefore choose a different approach. Unless explicitly stated otherwise, we will use the following set,
\begin{equation}
M=300\text{ MeV},\quad
m_\pi=140\text{ MeV},\quad
f_\pi=92.4\text{ MeV}.
\label{physinput}
\end{equation}
For illustration purposes, we tune the $\rho$-meson mass to an artificially low value, $m_\rho=500\text{ MeV}$. This results, by means of Eq.~\eqref{fixing}, in the parameter set
\begin{equation}
\begin{alignedat}{2}
G&=2.92\text{ GeV}^{-2},\qquad
&\Lambda&=817\text{ MeV},\\
G_V&=3.12\text{ GeV}^{-2},\qquad
&m&=3.30\text{ MeV}.
\end{alignedat}
\label{parameter_set}
\end{equation}


\subsection{Pion superfluid phase: condensates}
\label{subsec:NJLconds}

The presence of the isospin chemical potential $\muI$ explicitly breaks the $\gr{SU(2)}$ isospin symmetry down to its Abelian subgroup $\gr{U(1)}_{I_3}$, generated by $\tau_3$. When $\muI>m_\pi$, charged pions undergo BEC, which can be represented by nonzero expectation value of (one of) the $\pi^{1,2}$ fields. Such a condensate further breaks the remaining continuous isospin symmetry. At the same time, the pion condensate spontaneously breaks parity. However, there is a discrete $\gr{Z}_2$ symmetry which remains intact, generated by simultaneous parity transformation and a $\gr{U(1)}_{I_3}$ rotation by $180$ degrees, $e^{\imag\pi\frac{\tau_3}2}=\imag\tau_3$. Although we cannot prove rigorously that such \emph{modified parity} is not spontaneously broken in the pion superfluid phase, we take this as a plausible starting point~\footnote{We treat the pion condensate as the \emph{primary} condensate that triggers the phase transition. Other, secondary or induced, condensates may appear as a consequence, as long as they are consistent with the unbroken symmetry.}.

The unbroken symmetries of the pion superfluid phase consist of spacetime translations, spatial rotations and the modified parity. Apart from $\langle\pi^{1,2}\rangle$, condensates of the temporal vector fields are also consistent with these symmetries and therefore have to be included in our analysis, namely the neutral vector meson condensate $\langle\rho^3_0\rangle$ and the charged axial vector meson condensate $\langle a^{1,2}_0\rangle$.

From the effective action~\eqref{NJLeffaction}, we infer the mean-field thermodynamic potential density $\Omega$. Remarkably, even in presence of the pion and vector condensates, the determinant of the Dirac operator~\eqref{diracoperator} can still be factorized analytically in terms of quasiparticle dispersion relations $E^\pm_{\vek k}$, given by
\begin{equation}
\begin{split}
(E^\pm_{\vek k})^2={}&\vek k^2+M^2+\tilde\mu^2+\vec\pi^2+\vec a^2\\
&\pm2\sqrt{(M\tilde\mu-\vec\pi\times\vec a)^2+\vek k^2(\tilde\mu^2+\vec a^2)}.
\end{split}
\label{NJLdisprel}
\end{equation}
Here $\vec \pi\times\vec a$ is understood as $\pi^1a_0^2-\pi^2 a_0^1$ and $\tilde\mu\equiv\muI/2+\rho$, where $\rho$ denotes the vector mean field $\rho^3_0$. Likewise, the Lorentz index on the axial vector mean field $\vec a_0$ is dropped and it is denoted simply by $\vec a$, in addition to the pion mean field $\vec\pi$. The mean-field thermodynamic potential density then acquires the usual form, consisting of a condensate contribution and a fermionic quasiparticle contribution,
\begin{equation}
\begin{split}
\frac\Omega V={}&\frac{\sigma^2+\vec\pi^2}{4G}-\frac{\vec\rho^2+\vec a^2}{4G_V}-2N_c\sum_{s=\pm}\int\frac{\dd^3\vek k}{(2\pi)^3}\\
&\times\Bigl[E^s_{\vek k}+2T\log\bigl(1+e^{-\beta E^s_{\vek k}}\bigr)\Bigr].
\end{split}
\label{NJLTDpot}
\end{equation}

How do we in practice determine the expectation values of the mean fields? A straightforward solution would be to simultaneously solve a set of gap equations, obtained by differentiating the thermodynamic potential with respect to the mean fields. However, this may not be the most convenient way in case there are several solutions; one needs to compare their free energies in order to determine the actual equilibrium state. A common alternative is to look for the absolute minimum of the thermodynamic potential. Here we have to exercise some care though. The vector mean fields effectively play the role of Lagrange multipliers for the corresponding charge densities, and the thermodynamic potential is thus a negative definite function of their deviations from equilibrium. (Its second derivatives measure, up to a sign, the fluctuations of the charge densities.) The way out is to first solve the set of gap equations for $\vec\rho$ and $\vec a$, treating them as \emph{constraints} on the charge densities. Once the solution is plugged back  to $\Omega$, it can be subsequently minimized with respect to $\sigma$ and $\vec\pi$ as usual.

The analysis is further simplified by noting that the axial vector condensate is necessarily orthogonal to the pion condensate in the isospin space. This is easy to understand within $\chi$PT (see Appendix~\ref{app:ChPT}), but can also be proven directly within the NJL model (see Appendix~\ref{app:OGnality} for details). In addition, we can always use the symmetry to choose the orientation of the pion condensate at will. Finding thermodynamic equilibrium therefore boils down to determining the values of four mean fields, chosen without loss of generality as $\sigma$, $\pi^1$, $\rho^3_0$ and $a^2_0$.

\begin{figure}
\includegraphics[width=\columnwidth]{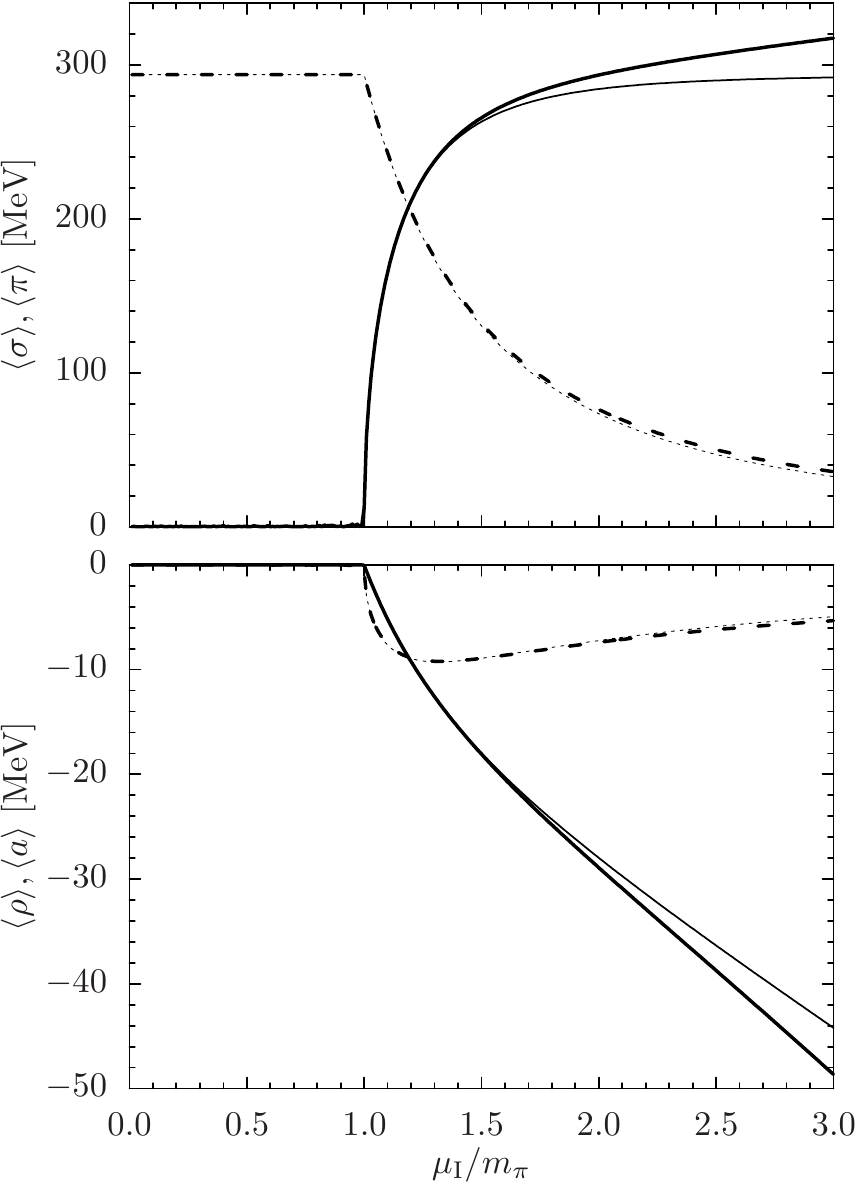}
\caption{The values of various condensates at $T=0$ as a function of the isospin chemical potential. Top panel: $\sigma$ (dashed) and $\pi$ (solid). Bottom panel: $\rho$ (solid) and $a$ (dashed). For comparison, we show the results both of the NJL model (thick lines) and of $\chi$PT (thin lines). The dimension-one condensates  $\rho$ and $a$ are related to the $\chi$PT prediction~\eqref{CHPTcondvect} by a factor of $4G_V$. The numerical results were obtained with the parameter set~\eqref{parameter_set}.}
\label{fig:condensates}
\end{figure}

The numerical results for the condensates at zero temperature as a function of $\muI$ are shown in Fig.~\ref{fig:condensates}. While the results for the (pseudo)scalar condensates are hardly surprising~\cite{He:2005nk}, we would like to emphasize the perfect agreement with $\chi$PT regarding the results for the vector condensates, see Eq.~\eqref{CHPTcondvect} in Appendix~\ref{app:ChPT}.  The predictions of $\chi$PT and their comparison to the NJL model do not involve any free parameters, and the agreement therefore confirms the consistency of our approach, including the mixing of pions with axial vector mesons.


\subsection{Pion superfluid phase: mass spectrum}
\label{subsec:NJLmasses}

There are two types of excitations that our mean-field analysis allows us to address: elementary fermionic (quark) and collective bosonic (meson). The dispersion relation of the fermionic quasiparticles is fully determined by the values of the condensates and Eq.~\eqref{NJLdisprel}. On the microscopic level, the formation of the pion condensate corresponds to pairing of $u$-quarks and $d$-antiquarks, at least at high isospin density where a Fermi sea of quarks is formed. It is therefore instructive to find the corresponding gap in the quasiparticle dispersion relation, $\Delta$~\footnote{In the BEC phase just after the onset of pion condensation, there is no Fermi sea and the quark quasiparticle dispersion relation has an absolute minimum at $\vek k=\vek0$. In this case, $\Delta=E^-_{\vek k=\vek0}$.}. This is determined by minimization of $E^-_{\vek k}$ with respect to momentum, and a short calculation yields
\begin{equation}
\Delta=\frac{\bigl|\tilde\mu|\vec\pi|-M|\vec a|\bigr|}{\sqrt{\tilde\mu^2+\vec a^2}}.
\label{NJLgap}
\end{equation}
Obviously, due to the presence of the axial vector condensate $\vec a$, the quasiparticle gap is \emph{not} given by the magnitude of the pion condensate $\vec\pi$ as usual. To assess the difference of $\Delta$ and $|\vec\pi|$ more quantitatively, we note that for our parameter set, the vector condensates are numerically much smaller than the chemical potential, hence to first order in $\vec a$ and $\vec\rho$ we can write
\begin{equation}
\Delta\approx|\vec\pi|\left(1-\frac{M|\vec a|}{\mu|\vec\pi|}\right).
\end{equation}
Using the $\chi$PT prediction for the condensates, Eqs.~\eqref{CHPTcond} and~\eqref{CHPTcondvect}, we then get the approximate expression
\begin{equation}
\Delta\approx|\vec\pi|\left(1-\frac{8G_Vf_\pi^2}{x^4}\right),
\end{equation}
where $x\equiv\muI/m_\pi$. For the parameter set~\eqref{parameter_set}, we have $8G_Vf_\pi^2\approx0.21$. The crossover from the BEC phase to the Fermi sea of quarks occurs roughly at $x\approx1.6$~\footnote{Here we define the position of the crossover as such $\muI$ that the in-medium constituent $u$-quark mass, $M-\muI/2$, drops to zero~\cite{He:2006tn,*Sun:2007fc}.}, and at this point the fermionic quasiparticle gap is reduced just by about $3\%$ compared to the pion condensate $\vec\pi$. Moreover, the two rapidly converge to each other as the chemical potential further increases.

Next we focus on the meson fluctuations. Their spectrum is again determined by Eq.~\eqref{NJLpolarization}. However, the functional derivatives are now to be taken in the equilibrium state that features all the condensates discussed in the previous subsection, $\langle\sigma\rangle$, $\langle\pi^1\rangle$, $\langle\rho^3_0\rangle$ and $\langle a^2_0\rangle$. The explicit expression for the isospin-singlet polarization function~\eqref{chiisosinglet} then has to be appropriately generalized,
\begin{equation}
\begin{split}
\chi^{(\mathcal{AB})}(p)={}&2G^{(\mathcal{AB})}+\imag N_c\int\frac{\dd^4k}{(2\pi)^4}\\
&\times\tr_{D,f}\bigl[\Gamma^{(\mathcal A)}S(k+p)\Gamma^{(\mathcal B)}S(k)\bigr],
\end{split}
\label{chisuperfluid}
\end{equation}
where the trace is now to be taken over the Dirac and flavor (isospin) space. The fermion propagator entering the polarization function includes the chemical potential and all the condensates,
\begin{equation}
\begin{split}
S(k)^{-1}\equiv{}&\slashed k+\frac12\muI\tau_3\gamma_0-m-\langle\sigma\rangle-\imag\langle\pi^1\rangle\tau_1\gamma_5\\
&+\langle\rho^3_0\rangle\tau_3\gamma_0+\langle a^2_0\rangle\tau_2\gamma_0\gamma_5.
\end{split}
\end{equation}

\begin{figure}
\includegraphics[width=\columnwidth]{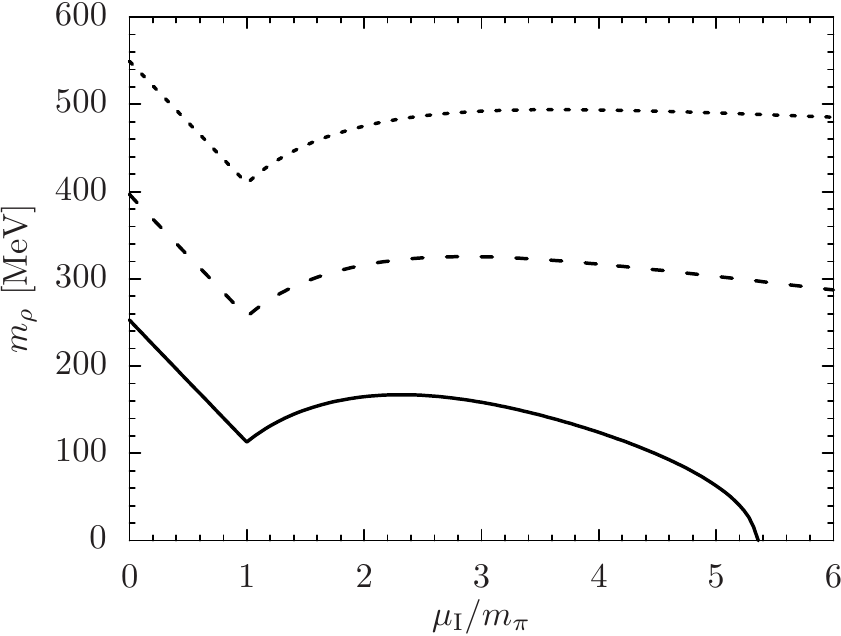}
\caption{Dependence of the in-medium charged $\rho$-meson mass on the isospin chemical potential. The solid, dashed and dotted line corresponds respectively to the vacuum $\rho$-meson mass of $250\text{ MeV}$, $400\text{ MeV}$, and $550\text{ MeV}$. In each case, the parameters of the model were adjusted in order to keep the other physical variables, listed in Eq.~\eqref{physinput}, fixed.}
\label{fig:rhomass}
\end{figure}

As in the vacuum phase, the polarization function will take a block-diagonal form, the structure of the blocks being determined by the unbroken symmetry in the equilibrium. The 22 physical degrees of freedom contained in the fields $\sigma$, $\vec\pi$, $\vec\rho_\mu$ and $\vec a_\mu$ then split into the following sectors, differing by their transformation properties under spatial rotations and the modified parity:
\begin{itemize}
\item The $\{\sigma,\pi^{1,2},\rho^3_0,a^{1,2}_0\}$ sector: 3 propagating scalar modes, one of which is the NG boson of the spontaneously broken isospin $\gr{U(1)}_{I_3}$ invariance.
\item The $\{\pi^3,\rho^{1,2}_0,a^3_0\}$ sector: 1 propagating pseudo\-scalar mode of mass $\muI$, corresponding to the massive NG boson of the  isospin symmetry~\cite{Watanabe:2013uya,Ebert:2005wr}.
\item The $\{\rho^3_i,a^{1,2}_i\}$ sector: 9 propagating vector modes.
\item The $\{\rho^{1,2}_i,a^3_i\}$ sector: 9 propagating axial vector modes.
\end{itemize}
The detailed mass spectrum of the meson excitations has to be determined numerically by finding the zeros of the determinant of the polarization matrix in each sector.

\begin{figure}
\includegraphics[width=\columnwidth]{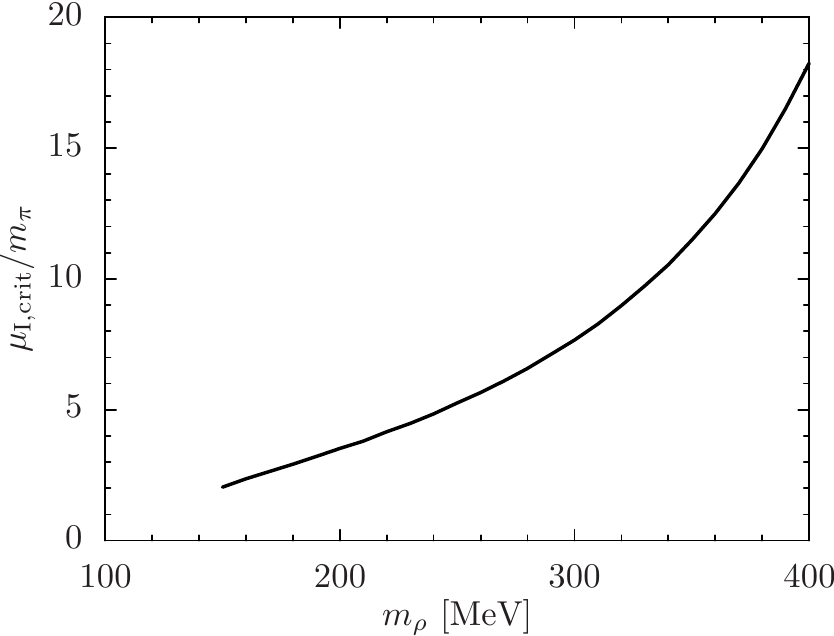}
\caption{Dependence of the critical isospin chemical potential, at which the mass of the lightest vector degree of freedom drops to zero, on the vacuum $\rho$-meson mass. The other physical  input variables were fixed according to Eq.~\eqref{physinput}.}
\label{fig:criticalmu}
\end{figure}

While it is in principle no problem to determine the full meson spectrum numerically, here we focus on the modes that are of particular interest to us, that is, pions and $\rho$-mesons. First of all, as already said above, one of the charged pions becomes exactly massless in the pion superfluid phase as a consequence of the Goldstone theorem. In addition, the neutral pion acquires a mass equal to $\muI$, that is, it is the massive NG boson of the isospin $\gr{SU(2)}$ symmetry~\cite{Watanabe:2013uya}. Both of these are \emph{exact} properties of the pion superfluid phase, guaranteed by the isospin symmetry, which are moreover easy to verify numerically.

Finally, we focus on the lightest vector degree of freedom, which corresponds to a mixture of the two charged $\rho$-meson fields. The dependence of its in-medium mass on the chemical potential is shown in Fig.~\ref{fig:rhomass}; the three curves correspond to different choices of the vacuum mass $m_\rho$ while keeping the other physical variables~\eqref{physinput} fixed. The figure reveals two characteristic properties of the vector mass spectrum.

First of all, just after the onset of pion BEC, the mass increases as a function of $\muI$. Note that the same behavior was predicted using effective field theory for charged vector mesons in two-color QCD~\cite{Harada:2010vy}, although the evidence from direct lattice simulations of the same theory seems somewhat inconclusive so far~\cite{Hands:2007uc}. Second, at sufficiently high chemical potential, a maximum occurs and the mass starts dropping again. Provided that the $\rho$-meson is sufficiently light in the vacuum, the in-medium mass drops to zero at some $\muI$ within the range of validity of the NJL model. This is a signature of $\rho$-meson BEC.

To appreciate how high $\muI$ has to be for us to observe $\rho$-meson BEC, we plot in Fig.~\ref{fig:criticalmu} the value of such ``critical'' $\muI$ as a function of the vacuum mass $m_\rho$. In QCD, the $\rho$-meson mass cannot be tuned freely. The parameter space of the NJL model is larger than that of QCD though. The way to think about Fig.~\ref{fig:criticalmu} is as follows. We start in a region of the NJL model parameter space where the onset of $\rho$-meson BEC is under theoretical control. Then we tune $m_\rho$, trying to extrapolate towards the physical surface within the parameter space, which matches QCD. It is obvious that such an extrapolation cannot be done reliably since already for values of $m_\rho$ shown in Fig.~\ref{fig:criticalmu}, the critical chemical potential lies way beyond the NJL model cutoff. However, we \emph{can} conclude with confidence that should $\rho$-meson BEC occur in QCD at all, it has to set in at a value of $\muI$ much higher than the vacuum $\rho$-meson mass.


\section{Linear sigma model analysis}
\label{sec:lsm}

In order to get some insight in the results obtained in the previous section, we now turn to a much simpler description of the meson spectrum, based on the linear sigma model. While this leads to a further extension of the parameter space, it has the great advantage of incorporating the physical degrees of freedom as elementary excitations rather than bound states of quarks. Since the mean-field approximation in the NJL model resums all quark loops but treats the bosonic fields as classical, the tree-level approximation is the appropriate equivalent in the linear sigma model.

In order to keep our discussion as simple as possible, we only include in the model those modes which are of most interest to us, that is, pions and $\rho$-mesons. This is reasonable since we expect that at sufficiently high chemical potentials, (some of) these will be the lightest degrees of freedom. Therefore, the model discussed below can be thought of as a low-energy effective theory for the lightest modes in the pion superfluid phase. As such, it cannot possess full chiral symmetry, not in a linear realization at least. We will instead impose the isospin $\gr{SU(2)}$ symmetry alone.

To that end, we note that the determinant of the Dirac operator~\eqref{diracoperator} can be made invariant under \emph{gauged} isospin $\gr{SU(2)}$ transformations, provided that the meson fields are assigned suitable transformation rules. In this case, the $\vec\rho_\mu$ field plays the role of a gauge field. Noting that such gauge invariance is broken explicitly only by the quadratic, condensate part of the action~\eqref{NJLeffaction}, we consider the following linear sigma model,
\begin{equation}
\La=\frac12(D_\mu\vec\pi)^2-\frac12m_\pi^2\vec\pi^2-\frac\lambda4(\vec\pi^2)^2-\frac14\vec F_{\mu\nu}\cdot\vec F^{\mu\nu}+\frac12m_\rho^2\vec\rho_\mu\cdot\vec\rho^\mu.
\label{lsm}
\end{equation}
Here $\vec F_{\mu\nu}\equiv\de_\mu\tilde{\vec\rho}_\nu-\de_\nu\tilde{\vec\rho}_\mu+g\tilde{\vec\rho}_\mu\times\tilde{\vec\rho}_\nu$ is the isospin $\gr{SU(2)}$ field-strength tensor and $D_\mu\vec\pi\equiv\de_\mu\vec\pi+g\tilde{\vec\rho}_\mu\times\vec\pi$ the covariant derivative of the pion field. Again motivated by the form of the Dirac operator~\eqref{diracoperator}, the vector meson field is shifted inside the gauge-invariant part of the Lagrangian according to
\begin{equation}
g\tilde{\vec\rho}_\mu\equiv (g\rho^1_\mu,g\rho^2_\mu,g\rho^3_\mu+\delta_{\mu0}\muI).
\end{equation}
The shift does not affect the $\rho$-meson mass term in Eq.~\eqref{lsm}. Altogether, the linear sigma model~\eqref{lsm} can be thought of as having been obtained from the NJL model by integrating out quarks, dropping the scalar and axial vector degrees of freedom, and truncating the Lagrangian to operators of dimension four or less. To what extent the parameters of the linear sigma model can actually be determined by this procedure is discussed in Appendix~\ref{app:matching}. Here we will treat $g$ and $\lambda$ as free parameters.


\subsection{Ground state}
\label{subsec:lsmground}

From our discussion of the NJL model, we expect that once the isospin chemical potential $\muI$ exceeds the pion mass $m_\pi$, a pion condensate appears, accompanied by a temporal vector condensate. The orientation of the pion condensate can be chosen at will; in this section we will use the following notation to distinguish the condensates from the dynamical fields,
\begin{equation}
\pi_0\equiv\langle\pi^1\rangle,\qquad
\rho_0\equiv\langle\rho^3_0\rangle.
\end{equation}
The condensates are determined by extremization of the static part of the Lagrangian~\eqref{lsm}, which leads to a set of two coupled gap equations,
\begin{equation}
\begin{split}
\lambda\pi_0^2-(\muI+g\rho_0)^2+m_\pi^2&=0,\\
g(\muI+g\rho_0)\pi_0^2+m_\rho^2\rho_0&=0.
\end{split}
\label{lsmgapeq}
\end{equation}
An analytic solution of this set of equations would require solving a cubic equation. We choose to gain insight by examining various special cases rather than by writing down a fully general solution.

First, the special case $g=0$ is well known. Namely, the pion and $\rho$-meson sectors decouple in this limit. Consequently, there is no vector condensate, while the pion condensate acquires the usual value,
\begin{equation}
\pi_0=\sqrt{\frac{\muI^2-m_\pi^2}\lambda}.
\label{g0case}
\end{equation}
Now we switch on the coupling $g$ but focus on the limit of very high chemical potential $\muI$. Then we can neglect the mass term in the first of the gap equations~\eqref{lsmgapeq}, which results in a set of equations that is easy to solve by a series expansion,
\begin{equation}
\begin{split}
\pi_0&=\left(\frac{\muI m_\rho^2}{g^2\sqrt\lambda}\right)^{1/3}+\mathcal O(\muI^{-1/3}),\\
\rho_0&=-\frac\muI g+\frac1g\left(\frac{\lambda\muI m_\rho^2}{g^2}\right)^{1/3}+\mathcal O(\muI^{-1/3}).
\end{split}
\label{lsmconds}
\end{equation}
The asymptotic behavior of the pion condensate as a function of $\muI$ is very different in the cases where $g$ is zero and nonzero. However, it is interesting to note that the difference disappears once the chemical potential is traded for the isospin density $\nI$, defined by the derivative of the static part of the Lagrangian~\eqref{lsm} with respect to $\muI$. Namely, both Eq.~\eqref{g0case} and Eq.~\eqref{lsmconds} reduce to
\begin{equation}
\pi_0\approx\left(\frac{\nI^2}\lambda\right)^{1/6}
\end{equation}
to the leading order in powers of $\nI$.


\subsection{Meson spectrum}
\label{subesc:lsmspectrum}

The meson dispersion relations can be determined by shifting the fields by the above-found condensates and subsequently expanding the Lagrangian to the second order in the field fluctuations. This is a completely routine procedure and we therefore omit details, merely providing a summary of the results. For the sake of simplicity we set the spatial momentum to zero and just overview the mass spectrum.
\begin{itemize}
\item The neutral $\pi$ sector: the mass of the neutral pion is found to be $\muI$, in accord with the prediction based on the exact isospin symmetry.
\item The charged $\pi$ sector: we find one gapless mode as expected, and one gapped mode with the mass
\begin{equation}
m_{\pi^-}=\sqrt{6(\muI+g\rho_0)^2-2m_\pi^2},
\end{equation}
where $\rho_0$ is determined implicitly by Eq.~\eqref{lsmgapeq}.
\item The neutral $\rho$ sector: the mass of neutral $\rho$ equals
\begin{equation}
m_{\rho^0}=\sqrt{m_\rho^2+g^2\pi_0^2},
\end{equation}
where $\pi_0$ is likewise determined by Eq.~\eqref{lsmgapeq}.
\item The charged $\rho$ sector: we find two gapped modes with masses given by
\begin{align}
\label{mrho}
m_{\rho^\pm}={}&\biggl[m_\rho^2+(\muI+g\rho_0)^2+\frac12g^2\pi_0^2\\
\notag
&\mp\sqrt{\frac14g^4\pi_0^4+2(\muI+g\rho_0)^2(2m_\rho^2+g^2\pi_0^2)}\biggr]^{1/2}.
\end{align}
\end{itemize}
We would now like to see to what extent this result can reproduce the behavior we found using the NJL model, see Fig.~\ref{fig:rhomass}. We first focus on the behavior of the lightest vector mode just after the onset of pion BEC. To that end, we define the nonrelativistic chemical potential for the charged pion as
\begin{equation}
\delta\muI\equiv\muI-m_\pi.
\end{equation}
Similarly to the asymptotic solution~\eqref{lsmconds} at very high $\muI$, it is now possible to solve the gap equations~\eqref{lsmgapeq} by a series expansion in $\delta\muI$. The result reads
\begin{equation}
\begin{split}
\pi_0&=\sqrt{\frac{2m_\pi\delta\muI}{\lambda+\frac{2g^2m_\pi^2}{m_\rho^2}}}+\mathcal O(\delta\muI^{3/2}),\\
\rho_0&=-\frac{\delta\muI}{1+\frac{\lambda m_\rho^2}{2g^2m_\pi^2}}+\mathcal O(\delta\muI^2).
\end{split}
\end{equation}
Inserting this back into Eq.~\eqref{mrho}, we find that just after the onset of pion BEC, the vector meson mass equals
\begin{equation}
m_{\rho^+}\approx m_\rho-m_\pi+\delta\muI\frac{\frac{g^2m_\pi}{2m_\rho}-\lambda}{\lambda+\frac{2g^2m_\pi^2}{m_\rho^2}}.
\end{equation}
We observe that the vector meson mass increases with~$\muI$ for $\muI\gtrsim m_\pi$, just as in Fig.~\ref{fig:rhomass}, provided that the gauge coupling $g$ is strong enough.

Next, we shift our attention to the region of high chemical potentials. It is not a priori obvious from Eq.~\eqref{mrho} whether or not the lightest mass $m_{\rho^+}$ drops to zero at sufficiently high $\muI$, which would signal BEC. However, it is easy to see that this indeed does happen when $\muI$ is so high that the following condition is satisfied~\footnote{This condition is equivalent to the requirement that the smaller of the two diagonal mass terms in the charged $\rho$ sector drops to zero.},
\begin{equation}
(\muI+g\rho_0)^2=m_\rho^2.
\end{equation}
Solving this together with the gap equations~\eqref{lsmgapeq}, we find that the mass of the lightest vector meson mode drops to zero for \emph{arbitrary} (finite and nonzero) values of the model parameters at a critical chemical potential determined by
\begin{equation}
\mu_\text{I,crit}=m_\rho+\frac{g^2}\lambda\left(m_\rho-\frac{m_\pi^2}{m_\rho}\right).
\label{muIcrit}
\end{equation}
This is not so surprising, for a nonzero expectation value of a non-Abelian charge density (in this case isospin) is known to act as a chemical potential for BEC of the associated gauge bosons~\cite{Gusynin:2003yu,*Watanabe:2014qla}.

Eq.~\eqref{muIcrit} makes it clear that the critical chemical potential is always larger than the vacuum mass $m_\rho$ (as long as $m_\rho>m_\pi$). Moreover, by tuning the ratio $g^2/\lambda$ suitably, it can in principle be made arbitrarily large. Regarding the dependence of the linear sigma model predictions on the unknown parameters $g$ and $\lambda$, it is useful to note that all the in-medium meson masses depend only on the vacuum masses $m_\pi$ and $m_\rho$, the chemical potential $\muI$, and the ratio $g^2/\lambda$. Together with Eq.~\eqref{muIcrit}, this implies that the dependence of the in-medium masses on $\muI$ within the linear sigma model is completely fixed once the values of $m_\pi$, $m_\rho$ and $\mu_\text{I,crit}$ are given.


\section{Conclusions and discussion}
\label{sec:conclusions}

We have investigated the possibility that vector mesons undergo BEC in QCD at sufficiently high isospin density and zero temperature. Using a combination of arguments based on the NJL and the linear sigma model, we concluded that condensation of $\rho$-mesons is disfavored by the preformed condensate of charged pions. As a consequence, if present at all, $\rho$-meson BEC is postponed to much higher chemical potentials than what would follow from a naive estimate based on the vacuum mass $m_\rho$.

We were able to determine the critical chemical potential for $\rho$-meson BEC for model parameter sets corresponding to artificially light $\rho$-mesons. Unfortunately, an extrapolation to the physical $\rho$-meson mass is not possible within our model setup since it leads far beyond the range of validity of the effective models. Ultimately, the question has to be settled by direct lattice simulation, which is possible at least in principle since QCD with nonzero isospin chemical potential does not suffer from the sign problem~\cite{Son:2000xc}.

Finally, let us make a comment regarding the stability of the $\rho$-mesons in the pion superfluid medium. While in the vacuum, $\rho$-mesons decay predominantly into two pions (as long as $m_\rho>2m_\pi$), the possible decay channels in the pion superfluid phase are constrained by the modified kinematics and by the modified parity invariance. As a consequence, a charged $\rho$-meson can decay into one neutral pion and one charged pion, but not into two charged pions. (The latter is \emph{not} forbidden by conservation of electric charge or isospin, which is spontaneously broken by the pion condensate.) Since the mass of the neutral pion equals $\muI$, it follows that at sufficiently high $\muI$, the lightest charged $\rho$-meson will be protected against decay into two pions by kinematics. Barring emergence of new decay channels in the superfluid medium, we can make the model-independent conclusion that the lightest charged $\rho$-meson will become stable, hence it makes sense to speak of its BEC in the first place.


\section*{Acknowledgments}

T.~B. would like to thank Denis Parganlija, Owe Philipsen, and Dirk H.~Rischke for helpful discussions at various stages of this project. X.-G.~H. thanks Gaoqing Cao, Lianyi He, and Swagato Mukherjee for useful discussions. We also appreciate correspondence with Yoshimasa Hidaka pertinent to rigorous constraints on $\rho$-meson condensation. X.-G.~H. is supported by the 1000 Young Talents Program of China and by NSFC with Grant No. 11535012.


\appendix

\section{Chiral perturbation theory treatment}
\label{app:ChPT}

In this appendix we briefly review how the condensates in the pion superfluid phase, discussed in Sec.~\ref{subsec:NJLconds}, can be addressed using $\chi$PT. The pseudo-NG boson degrees of freedom of $\chi$PT are encoded in a unitary matrix variable $\Sigma$, in terms of which the leading-order $\chi$PT Lagrangian reads~\cite{Son:2000xc}
\begin{equation}
\La=\frac{f_\pi^2}4\tr\bigl[D_\mu\he\Sigma D^\mu\Sigma+m_\pi^2(\Sigma+\he\Sigma)\bigr].
\label{LagCHPT}
\end{equation}
Under a chiral transformation, defined by the left- and right-handed unitary matrices $U_L$ and $U_R$, the field $\Sigma$ transforms as $\Sigma\to U_L\Sigma\he U_R$. Therefore, it couples to background matrix gauge fields $L_\mu$ and $R_\mu$ of the chiral group via the covariant derivative
\begin{equation}
D_\mu\Sigma\equiv\de_\mu\Sigma-\imag L_\mu\Sigma+\imag\Sigma R_\mu.
\end{equation}


\subsection{Ground state at nonzero isospin density}

Introducing the isospin chemical potential $\muI$ is equivalent to setting the background fields $L_\mu$ and $R_\mu$ to
\begin{equation}
L_\nu=R_\nu=\frac12\delta_{\nu0}\muI\tau_3.
\end{equation}
Likewise, the $2\times2$ unitary matrix $\Sigma$ can always be decomposed into a linear combination of the unit and Pauli matrices; the presence of the scalar and pseudoscalar condensates $\langle\sigma\rangle$ and $\langle\pi^1\rangle$ then corresponds to the parameterization
\begin{equation}
\Sigma=\cos\theta+\imag\tau_1\sin\theta
\end{equation}
with coordinate-independent angle $\theta$. This angle is found by minimization of the dimensionless potential
\begin{equation}
V(\theta)=-\frac{\La}{f_\pi^2m_\pi^2}=-\frac12x^2\sin^2\theta-\cos\theta,
\end{equation}
where we introduced the dimensionless parameter $x\equiv\muI/m_\pi$. A pion condensate appears for $x>1$ and is given implicitly by $\cos\theta=1/x^2$. This corresponds to the vacuum expectation value
\begin{equation}
\langle\Sigma\rangle=\openone\frac1{x^2}+\imag\tau_1\sqrt{1-\frac1{x^4}},
\label{CHPTcond}
\end{equation}
which determines the dependence of the scalar and pseudoscalar condensates on $\muI$ up to an overall factor.


\subsection{Vector and axial vector condensates}

The pion condensate, defining the superfluid ground state, induces secondary condensates, admitted by the unbroken symmetry. These are in particular the vector and axial vector condensate. Within $\chi$PT, they can be calculated by introducing additional \emph{infinitesimal} background fields $\vec V_\mu$ and $\vec A_\mu$ such that
\begin{equation}
L_\mu=\frac12(\vec V_\mu-\vec A_\mu)\cdot\vec\tau,\qquad
R_\mu=\frac12(\vec V_\mu+\vec A_\mu)\cdot\vec\tau.
\end{equation}
The condensates can be obtained by taking a derivative of the Lagrangian with respect to the backgrounds fields, evaluated in the ground state. A straightforward manipulation leads to
\begin{equation}
\begin{split}
\langle\bar\psi\tfrac{\tau_a}2\gamma^0\psi\rangle&\equiv\frac{\de\La}{\de V^a_0}=\delta_{a3}m_\pi f_\pi^2x\left(1-\frac1{x^4}\right),\\
\langle\bar\psi\tfrac{\tau_a}2\gamma^0\gamma_5\psi\rangle&\equiv\frac{\de\La}{\de A^a_0}=\delta_{a2}m_\pi f_\pi^2\frac1{x}\sqrt{1-\frac1{x^4}}.
\end{split}
\label{CHPTcondvect}
\end{equation}
These expressions completely fix the dependence of the vector and axial vector condensates in the pion superfluid phase on the chemical potential. In the vacuum phase ($x\leq1$) both condensates are zero.


\section{Orthogonality of the pion and axial vector condensates}
\label{app:OGnality}

In Sec.~\ref{subsec:NJLconds} we claimed that the axial vector condensate is necessarily orthogonal to the pion condensate in isospin space. This fact was used to simplify the numerical calculation. In this appendix we give a proof of this claim within the NJL model. First, we write down the gap equations for $\sigma$ and $\vec\pi$ derived from the thermodynamic potential~\eqref{NJLTDpot},
\begin{widetext}
\begin{equation}
\begin{split}
\sigma-4 G N_c\sum_{s=\pm}\int\frac{\dd^3{\vek k}}{(2\pi)^3}\frac{1}{E_{\vek k}^s}\left[M+s\tilde{\mu} \frac{M\tilde{\mu}-{\vec \pi}\times{\vec a}}{\sqrt{(M\tilde{\mu}-{\vec \pi}\times{\vec a})^2+{\vek k}^2(\tilde{\mu}^2+{\vec a}^2)}}\right]\left[1-2f(E_{\vek k}^s)\right]&=0,\\
\pi^1-4 G N_c\sum_{s=\pm}\int\frac{\dd^3{\vek k}}{(2\pi)^3}\frac{1}{E_{\vek k}^s}\left[\pi^1-s a_0^2\frac{M\tilde{\mu}-{\vec \pi}\times{\vec a}}{\sqrt{(M\tilde{\mu}-{\vec \pi}\times{\vec a})^2+{\vek k}^2(\tilde{\mu}^2+{\vec a}^2)}}\right]\left[1-2f(E_{\vek k}^s)\right]&=0,\\
\pi^2-4 G N_c\sum_{s=\pm}\int\frac{\dd^3{\vek k}}{(2\pi)^3}\frac{1}{E_{\vek k}^s}\left[\pi^2+s a_0^1\frac{M\tilde{\mu}-{\vec \pi}\times{\vec a}}{\sqrt{(M\tilde{\mu}-{\vec \pi}\times{\vec a})^2+{\vek k}^2(\tilde{\mu}^2+{\vec a}^2)}}\right]\left[1-2f(E_{\vek k}^s)\right]&=0,
\end{split}
\label{appenb1}
\end{equation}
where $f(x)\equiv1/[1+\exp{(x/T)}]$ is the Fermi-Dirac distribution. If we introduce the abbreviations,
\begin{equation}
\begin{split}
x&\equiv1-4 G N_c\sum_{s=\pm}\int\frac{\dd^3{\vek k}}{(2\pi)^3}\frac{1}{E_{\vek k}^s}\left[1-2f(E_{\vek k}^s)\right],\\
y&\equiv4 G N_c\sum_{s=\pm}\int\frac{\dd^3{\vek k}}{(2\pi)^3}\frac{s}{E_{\vek k}^s}\frac{M\tilde{\mu}-{\vec \pi}\times{\vec a}}{\sqrt{(M\tilde{\mu}-{\vec \pi}\times{\vec a})^2+{\vek k}^2(\tilde{\mu}^2+{\vec a}^2)}}\left[1-2f(E_{\vek k}^s)\right],
\end{split}
\label{appenb2}
\end{equation}
\end{widetext}
the above gap equations can be cast as
\begin{equation}
\begin{split}
x M-m-y\tilde{\mu}&=0,\\
\begin{pmatrix}
\pi^1 & a_0^2 \\
\pi^2 & -a_0^1
\end{pmatrix}
\begin{pmatrix}
x \\ y
\end{pmatrix}&=0,
\end{split}
\label{appenb3}
\end{equation}
From the first line in Eq.~\eqref{appenb3} we can conclude that $x$ and $y$ cannot vanish simultaneously as long as the current quark mass $m$ is nonzero. Then the second line requires the determinant of the coefficient matrix to be zero which implies $\vec \pi\cdot\vec a=0$. This orthogonality of the pion and axial vector condensates in the isospin space is consistent with the expectation from the $\chi$PT.


\section{Matching the NJL and linear sigma models}
\label{app:matching}

The linear sigma model analyzed in Sec.~\ref{sec:lsm} was motivated by the NJL model, yet its couplings were treated as free parameters. In this appendix we wish to take a closer look at the extent to which the linear sigma model parameters can be \emph{determined} from the NJL model.

As far as the correlators of the meson interpolating fields are concerned, the bosonized action~\eqref{NJLeffaction} is equivalent to the original NJL model~\eqref{NJLlag}. The corresponding approximate linear sigma model can then be obtained by expanding the action in powers of the meson fields and their derivatives, and keeping only operators with canonical dimension four or less.

The derivative expansion of the logarithm of the Dirac operator $\De$ is most conveniently carried out using the method of covariant symbols~\cite{Pletnev:1998yu,*Salcedo:2000hp}. First we drop the scalar and axial vector fields in order to keep just the field content of the model analyzed in Sec.~\ref{sec:lsm}. Using the fact that the determinant of the Dirac operator $\De$ is gauge-invariant, we can actually keep the pion field only; the dependence on $\vec\rho_\mu$ can subsequently be restored by imposing the gauge invariance. This leads to the linear sigma model Lagrangian
\begin{equation}
\begin{split}
\La={}&d_1(D_\mu\vec\pi)^2+d_2\vec\pi^2+d_3(\vec\pi^2)^2\\
&+d_4\vec F_{\mu\nu}\cdot\vec F^{\mu\nu}+d_5\vec\rho_\mu\cdot\vec\rho^\mu,
\end{split}
\label{efflag}
\end{equation}
where the effective couplings are given by
\begin{equation}
\begin{gathered}
d_1=2N_cI_2,\quad
d_3=-2N_cI_2,\quad
d_5=\frac1{4G_V},\\
d_2=-\frac1{4G}+4N_cI_1,\quad
d_4=-\frac{N_c}6(I_2+\tfrac12M^2I_3).
\end{gathered}
\end{equation}
The coefficients $I_n$ stand for a set of Euclidean loop integrals, generalizing Eqs.~\eqref{tadpole} and~\eqref{NJLIPiIntegral},
\begin{equation}
I_n\equiv\int\frac{\dd^4k_\text{E}}{(2\pi)^4}\frac1{(k_\text{E}^2+M^2)^n}.
\label{Indef}
\end{equation}
Those with $n\leq2$ are ultraviolet divergent and thus require and explicit cutoff. Integration by parts shows that the integrals satisfy the recursive relation
\begin{equation}
I_{n+1}=\left(1-\frac2n\right)\frac{I_n}{M^2}.
\label{recursive}
\end{equation}
Once we know the coefficients in Eq.~\eqref{efflag} in terms of the NJL model couplings, it is a matter of a simple field redefinition to match this Lagrangian to that of Eq.~\eqref{lsm}, leading to the matching relations
\begin{equation}
\begin{alignedat}{3}
m_\pi^2&=-\frac{d_2}{d_1},\qquad
&m_\rho^2&=-\frac{d_5}{8d_4},\\
\lambda&=-\frac{d_3}{d_1^2},\qquad
&g&=\frac1{2\sqrt{-d_4}}.
\end{alignedat}
\label{matching}
\end{equation}
Unfortunately, it turns out that the predictions of such a matched linear sigma model can be far from those of the original NJL model already in the vacuum, especially regarding the $\rho$-meson mass. It is therefore hopeless to use such a model to gain a quantitative insight into the spectrum of mesons in the pion superfluid phase.

\begin{figure}
\includegraphics[width=\columnwidth]{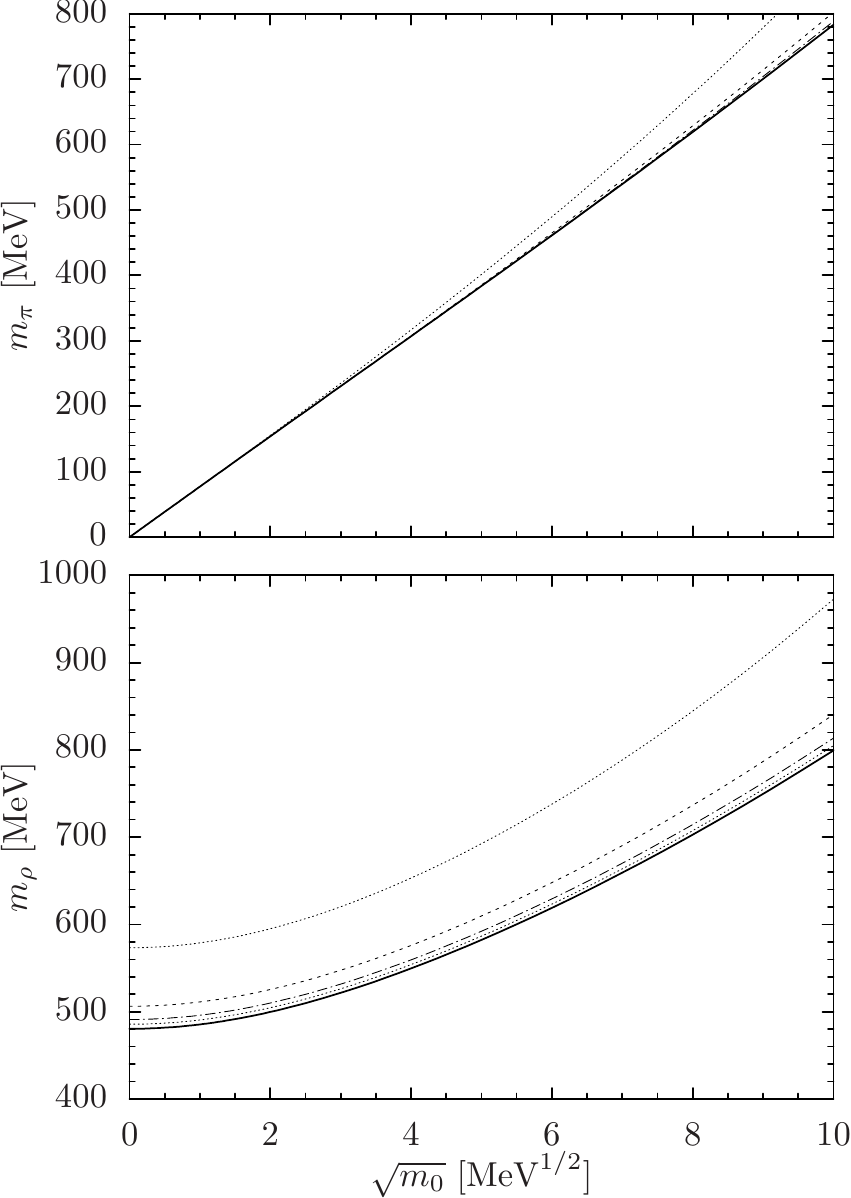}
\caption{The pion and $\rho$-meson vacuum mass as a function of the current quark mass. The solid lines denote the pole masses obtained from Eq.~\eqref{fixing}. The dotted, dashed, dash-dotted and dotted lines correspond to the approximate masses, obtained by expanding the pion and $\rho$-meson polarization functions to the first, second, third and fourth order in squared momentum, respectively. The numerical results were obtained with all model parameters but $m$ fixed according to Eq.~\eqref{parameter_set}.}
\label{fig:bosonization}
\end{figure}

The source of the problem is twofold. First, the couplings $d_i$ of the Lagrangian~\eqref{efflag} suffer from an ambiguity which arises from the ultraviolet divergence of the momentum integrals involved. This can be seen already from the fact that the coupling $d_4$ differs from what one would expect based on the transverse part of the $\rho$-meson polarization tensor~\eqref{NJLrhopolar}. As a matter of fact, naive application of the recursive relation~\eqref{recursive} would suggest that $I_3=0$, which would remove the discrepancy. Of course, the integral $I_3$ itself is ultraviolet-finite and by its definition~\eqref{Indef} nonzero and positive. This is not a problem of the method of covariant symbols or even of the effective Lagrangian~\eqref{efflag}. Namely the same ambiguity affects the meson polarization functions in the NJL model; it is well known that the ultraviolet cutoff is a part of the definition of the model.

The second problem is the truncation of the expansion in powers of derivatives, wherein only the leading, two-derivative kinetic term is kept. To study the effect of this truncation, we consider the pole conditions for the pion and $\rho$-meson mass, given in Eq.~\eqref{fixing}. Modulo the ambiguity just discussed, expanding the meson polarization functions to first order in squared momentum and solving for the pole should give the same mass spectrum as the linear sigma model~\eqref{efflag}. We can then subsequently improve upon the linear sigma model by adding higher order terms to the series expansion of the meson polarization functions. (This is equivalent to adding bilinear terms with four or more derivatives to the linear sigma model Lagrangian.)

The result of this procedure is shown in Fig.~\ref{fig:bosonization}. Obviously, the linear sigma model gives an accurate prediction for the pion mass unless the quarks are very heavy. Moreover, the upper plot in Fig.~\ref{fig:bosonization} confirms numerically the Gell-Mann-Oakes-Renner relation: the  squared pion mass is proportional to the current quark mass. However, the prediction for the $\rho$-meson mass is much worse. This is natural since the $\rho$-meson is heavier than the pion and thus the series expansion of its polarization tensor converges more slowly at the pole. While the values shown in the lower plot of Fig.~\ref{fig:bosonization} exhibit a still relatively acceptable error of roughly 20\%, for the physical $\rho$-meson mass the matched linear sigma model clearly cannot be taken quantitatively seriously. This is the reason why in the discussion in Sec.~\ref{sec:lsm}, we prefer to treat the couplings $\lambda$ and $g$ as free parameters; using Eq.~\eqref{matching} instead might give a false impression of having the linear sigma model predictions under numerical control.


\bibliography{references}


\end{document}